# Diagrammatization of the Transmission Control Protocol

**Sabah Al-Fedaghi**

**Computer Engineering Department, Kuwait University**
**P.O. Box 5969 Safat 13060 Kuwait**

**Abstract**

With the wide spread of Internet services, developers and users need a greater understanding of the technology of networking. Acquiring a clear understanding of communication protocols is an important step in understanding how a network functions; however, many protocols are complicated, and explaining them can be demanding. In addition, protocols are often explained in terms of traffic analysis and oriented toward technical staff and those already familiar with network protocols. This paper aims at proposing a diagrammatic methodology to represent protocols in general, with a focus on the Transmission Control Protocol and Secure Sockets Layer in particular. The purpose is to facilitate understanding of protocols for learning and communication purposes. The methodology is based on the notion of flow of "primitive" things in a system with six stages: creation, release, transfer, arrival, acceptance, and processing. Though the method presents a basic description of protocols without in-depth analysis of all aspects and mechanisms, the resultant conceptual description is a systematic specification that utilizes a few basic notions that assist in illustrating functionality and support comprehension.

***Keywords:*** *conceptual model, Transmission Control Protocol, Secure Sockets Layer, protocol specification, flowthing model.*

## 1. Introduction

With the increasing trend to sharing of hardware, resources, and data along with the wide spread of Internet services, developers and users need greater understanding of the technology of networking. Networking is a crucial and sensitive factor in effective usage of information technology. Because of the importance of this technology for making networking decisions, e.g., structural, operational, purchasing, management, …, a critical need has developed for understanding computer networks at all levels by users, managers, developers, designers, and others.

However, it is not required to know everything. Most users might never completely understand intricacies of networking and how its aspects such as security are organized; nevertheless, providing some basic knowledge can help immensely. This level of understanding can be supported best by conceptual descriptions.

Protocols play a crucial role in today's communication world. Communication protocols provide traffic control that facilitates communication among computers on a network. Acquiring a clear understanding of protocols is an important step in understanding how a network functions.

According to Lahdenmäki [1], "Many data communications protocols are complicated and explaining them can be demanding." Most current methods of explaining protocols focus on traffic analysis and are oriented toward technical staff and those already familiar with network protocols.

In this paper we focus on a certain level of the network stack. Protocols work as a layered communication system as in the TCP/IP model, which consists of the four layers *link* (device driver and interface card), *network* (e.g., IP protocols), *transport* (e.g., the TCP protocol), and *application* (includes FTP and DNS) [2, 3, 4, 5]. The transport protocol, and especially its representative TCP, is at the center of our attention. "It is important to understand TCP if one is to understand the historic, current and future architecture of the Internet protocols. Most applications on the Internet make use of TCP, relying upon its mechanisms that ensure safe delivery of data across an unreliable IP layer below" [6]. According to Lacković et al. [7],

> The concept of the transport protocol in most cases gives a student a vague picture that is difficult to comprehend. This is caused by a logical end-to-end service in a connectionless environment like Internet.

The most common method of describing TCP (and, in general, communication protocols) is to explain its mechanisms and characteristics verbally and in pictures or diagrams.

> This approach is usually characterized by the lack of comprehension. On the other hand almost every student has his own image of the Internet and its protocols acquired from his experience as an Internet user. This image has little or no theoretical background. [7]

Another reason to facilitate understanding of data communication protocols is the issue of network security. With the widespread use of the Internet, computer network security has become an important aspect of its operation [8]. An understanding of protocols in general, especially security protocol, is now a necessity when using the Internet.

This paper aims at proposing a diagrammatic methodology to represent protocols in general, and TCP and SSL in particular. It is based on the notion of flow of "primitive" things in a system with six stages: creation, release, transfer, arrival, acceptance, and processing. The resultant conceptual description is a systematic specification that utilizes a few basic notions that assist in illustrating functionality and support comprehension. The method represents a basic description of TCP without an in-depth analysis of all aspects and mechanisms of the protocol; nevertheless, it is suitable to use to describe TCP to any level of detail.

To make this paper self-contained, the materials in the following section are summarized from a series of papers that have applied the model in several application areas [9–11].

## 2. Flowthing Model

The Flowthing Model (FM) is a uniform method for representing things that flow, called flowthings. Flow in FM refers to the exclusive (i.e., being in one and only one) transformation among six states (also called stages) of transfer, process, create, release, arrive, and accept (see Fig. 1). All other states are not generic states. For example, we may have stored created flowthings, stored processed flowthings, stored received flowthings, etc. Flowthings can be released but not transferred (e.g., the channel is down), or arrived but not accepted, … We will use *Receive* as a combined stage of *Arrive* and *Accept* whenever appropriate, i.e., whenever arriving flowthings are always accepted. The fundamental elements of FM are as follows:

**Flowthing**: A thing (e.g., information, material, money, shuttle, good) that has the capability of being created, released, transferred, arrived, accepted, and processed while flowing within and between systems.

**A flow system** (referred to as flowsystem), as depicted in Fig. 1, comprises the internal flows of a system with the six stages and transactions among them.

**Spheres and subspheres**: Spheres and subspheres are the environments of the flowthing, such as a transistor, a battery, and a wire, which form the sphere of an electrical current, the flowthing.

**Triggering**: Triggering is a transformation (denoted by a dashed arrow) from one flow to another, e.g., flow of electricity triggers the flow of air.

## 3. Example

Many issues must be addressed in the area of communication protocols, especially the type of protocol used, its initialization, termination, the size of transmitted messages, errors, and damaged transmission. A simple example is the flow control in the so-called Stop-and-Wait protocol (see [12, 13]). It messages are called frames. It works as follows (see Fig. 2) [13]:

- The receiver sends an acknowledgment when a frame is received;
- Upon sending a frame, the sender waits for an acknowledgment, then sends another frame.

Fig. 3 shows the corresponding FM representation of this scenario.

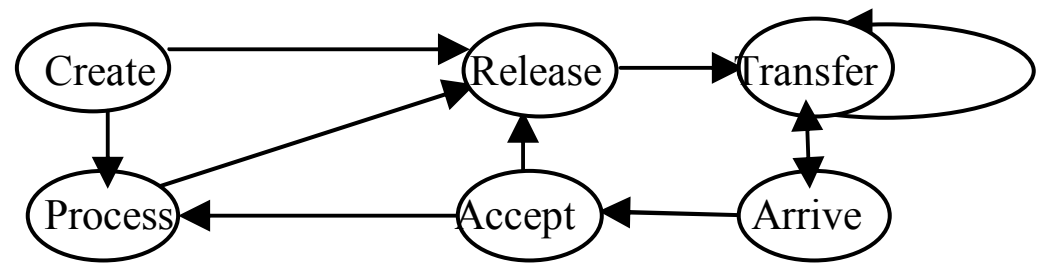

Fig. 1 Flowsystem

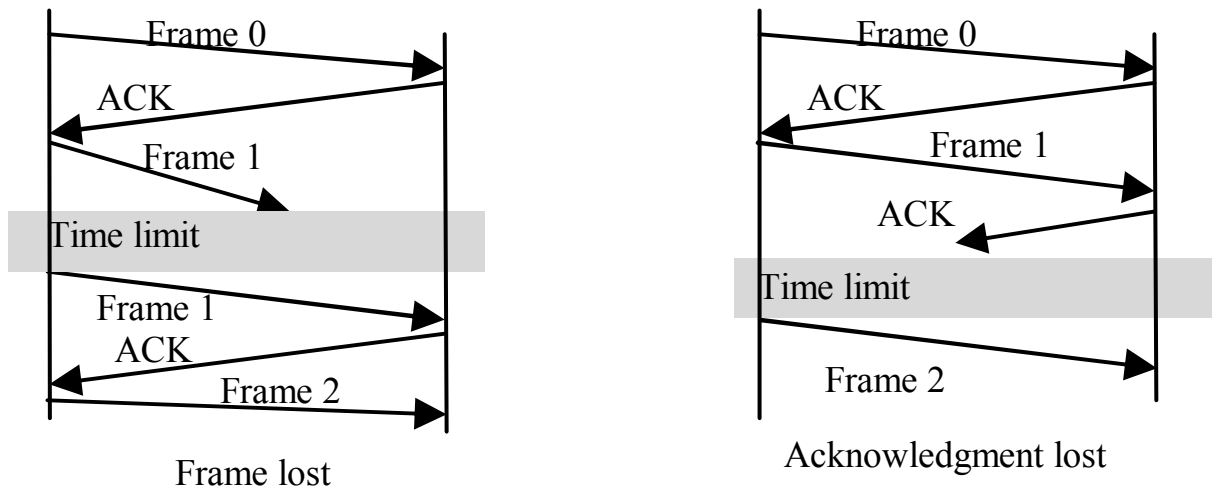

Fig. 2 Stop-and-Wait protocol

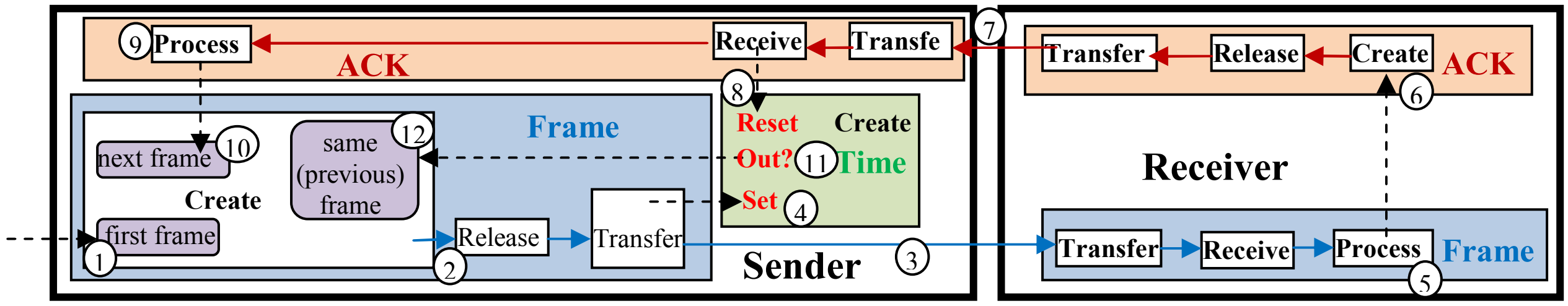

Fig. 3 FM representation of the Stop-and-Wait protocol

In Fig. 3, there are two spheres: Sender and Receiver. The sender has three subspheres: Frame, ACK, and Time. The receiver has two subspheres: Frame, and ACK. First, the sender is triggered to create (circle 1 in the figure) a first frame that is released (2) and flows to the receiver (3). Transferring a frame from the sender triggers setting the waiting time (4) for an acknowledgment. In the frame flowsystem of the receiver, the arriving frame is processed (5) and triggers the creation of ACK (6). Notice that, for simplicity's sake, when a sphere/subsphere (e.g., Frame subsphere of the receiver) has a single flowsystem (Frame flowsystem), we draw them in one rectangle.

ACK flows to the sender (7), where its arrival resets the wait time (8). It is processed (9) to trigger the creation of the next frame that flows to the receiver as previously. If time runs out (11) before the arrival of ACK, this triggers the creation and sending of the same frame again (12).

Note how the FM representation depicts a continuous sequence of acts involved in communication instead of the "two solid walls" of the sequence diagram shown in Fig. 2. Its representation is characterized by continuity of different threads, making it possible to have a tight series of superimposed protocol rules. Fig. 2 does not present a complete picture, and the specification is fragmented (triggering events behind the lines) and has vague semantics (e.g., half an arrow to represent incomplete communication).

In addition, the FM description provides a base for superimposing coordination with other tools such as synchronization, security constraints, and logical operations. For example, it is clear that the possibility exists of "premature time-out", that is, triggering sending of the same frame again (12), and during this, ACK arrives at 7. However, the description exposes the internal stages of operations of the sender and receiver; thus, it is possible to develop several alternative solutions to such a problems, such as coordinating the release (2) of a frame for the second time with the latest arrival of a late ACK, as illustrated in Fig. 4.

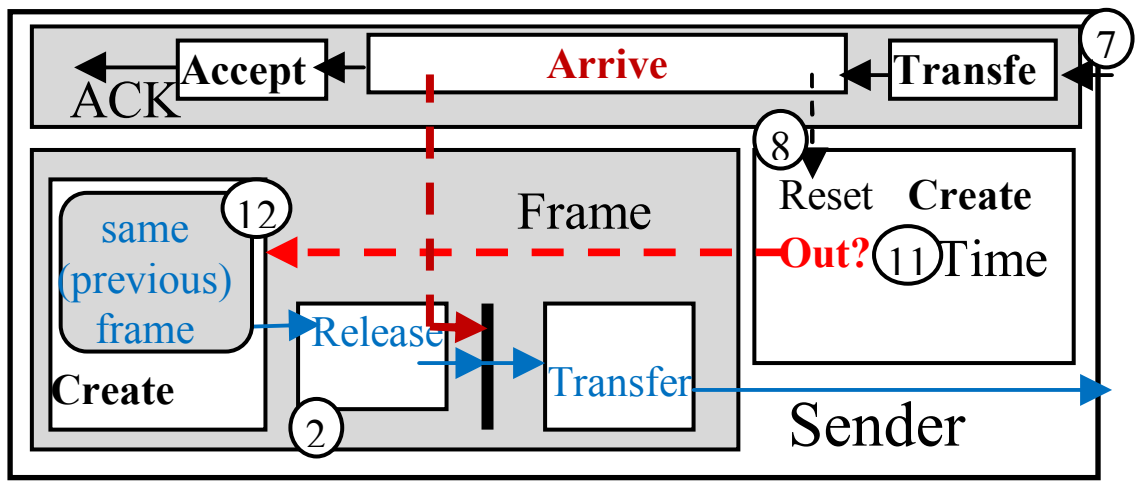

Fig. 4 Synchronization between generation of the same frame again and arrival of late ACK.

## 4. Transmission Control Protocol

The Transmission Control Protocol (TCP) [14] is the most common transport layer used to ensure that data packets are delivered in a reliable manner from one computer to another. Its importance in the overall network architecture comes from its role as a vehicle to relieve the application from communication details in the lower layers, and to facilitate data transportation across the network. In addition, it is one of the two original components of the Internet protocol suite, complementing the Internet Protocol (IP), and therefore the whole suite is referred to as TCP/IP. TCP/IP maps to a four-layer conceptual model: Application, Transport, Internet, and Network Interface. The Transport Layer facilitates data flow between two hosts where TCP is used, and a reliable connection is required. Reliability, here, denotes that the sender always knows whether a packet has reached the receiver through an acknowledgment of the arrival of the packet at its destination; otherwise the sender resends the packet. The sender also varies the rate of sending packets according to traffic congestion. Ports are used to conduct multiple simultaneous processes on one host by providing additional addresses to route information. A port and an IP address together form a socket.

TCP supports a full-duplex (simultaneous) connection with two byte-streams, one for each direction. To establish a connection between two hosts X to Y:
- X sends to Y a *SYN* (Synchronize - initiates a connection) packet with a randomly generated sequence number. If X does not receive an acknowledgment within a certain specified time, it resends the packet.
- *Remote host* Y sends back a *SYN+ACK* packet containing the next sequence number from X.
- X responds with an ACK packet (Acknowledges received data) with its acknowledgment number.

The sequence number and acknowledgment number fields are used to keep track of the byte count in the data streams. Both ends use their own, independent sequence numbers, and acknowledgment is related to the number of bytes transferred. Flow control is accomplished through the receiver's ability to control the size of the segment dispatched by the sender by using the *Window* field (the maximum number of bytes that can be accepted) of an acknowledgment packet.

Actually, TCP assigns more information—e.g., source port number, destination port number—to the data coming from the upper layer for use in ensuring communication reliability. A number of flags (1-bit Boolean fields) in the TCP header are used to control the state of a connection.

## 5. Description of TCP

To demonstrate the expressive power of FM, we utilize a specific description of the sequence of events in a TCP given by Kak [15]. In the paragraphs in italics throughout this section, we summarize Kak's description [15], leaving out irrelevant details, and sometimes copying some of his sentences. After each paragraph we will show the (almost) corresponding FM representation. Starting with construction of a TCP segment, the sequence of communication events proceeds as follows.

*Bytes are grouped together to form a TCP segment (datagram, packet) that consists of a header (with initial sequence number of client) and the data. The TCP segments are passed on to the IP layer for transmission.*

Fig. 5 is an approximate FM representation of the TCP protocol.

In general, TCP is formalized as the state diagram given in Fig. 6. This state diagram has been extensively discussed in protocol engineering. We include it here for the purpose of, superficially, contrasting the two methodologies of diagramming.

In Fig. 5, *data* flows from Application (circle 1), is processed (2), and triggers (3)—in synchronization with a created *TCP header* (4)—the creation of a *TCP segment* (5) that flows to the IP (6). Note that data, the TCP header, and the TCP segment are flowthings, each with its own flowsystem. The processing of data (2) and the creation of a TCP header (4) trigger the creation of a TCP segment (5).

*The decision how to break the byte stream into TCP segments depends on the Window field sent by the receiver. The receiver TCP sets a value for this field depending on the amount of memory allocated to the connection for the purpose of buffering the received data. This is referred to as the TCP's sliding window algorithm for flow control.*

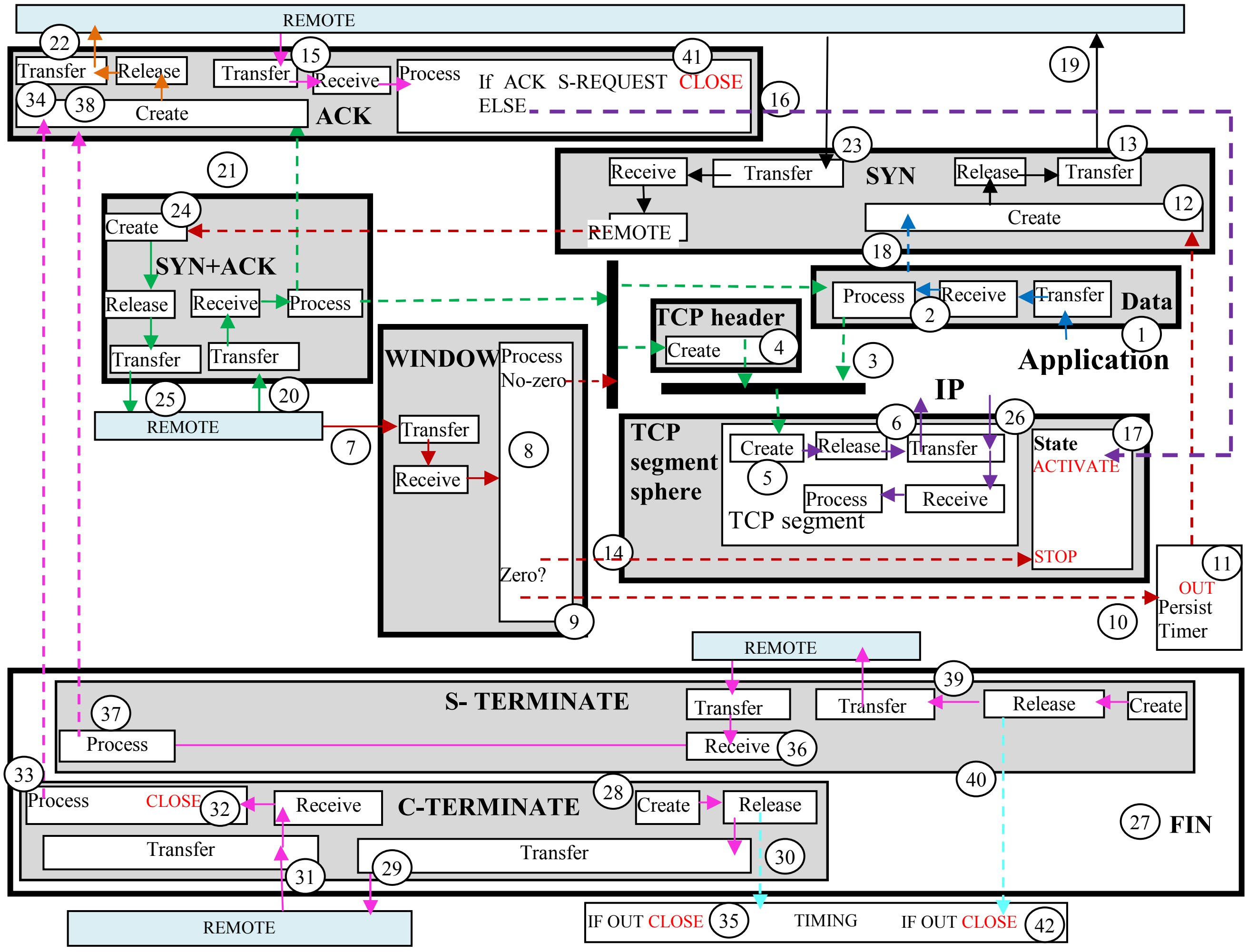

Fig. 5 The FM specification of TCP

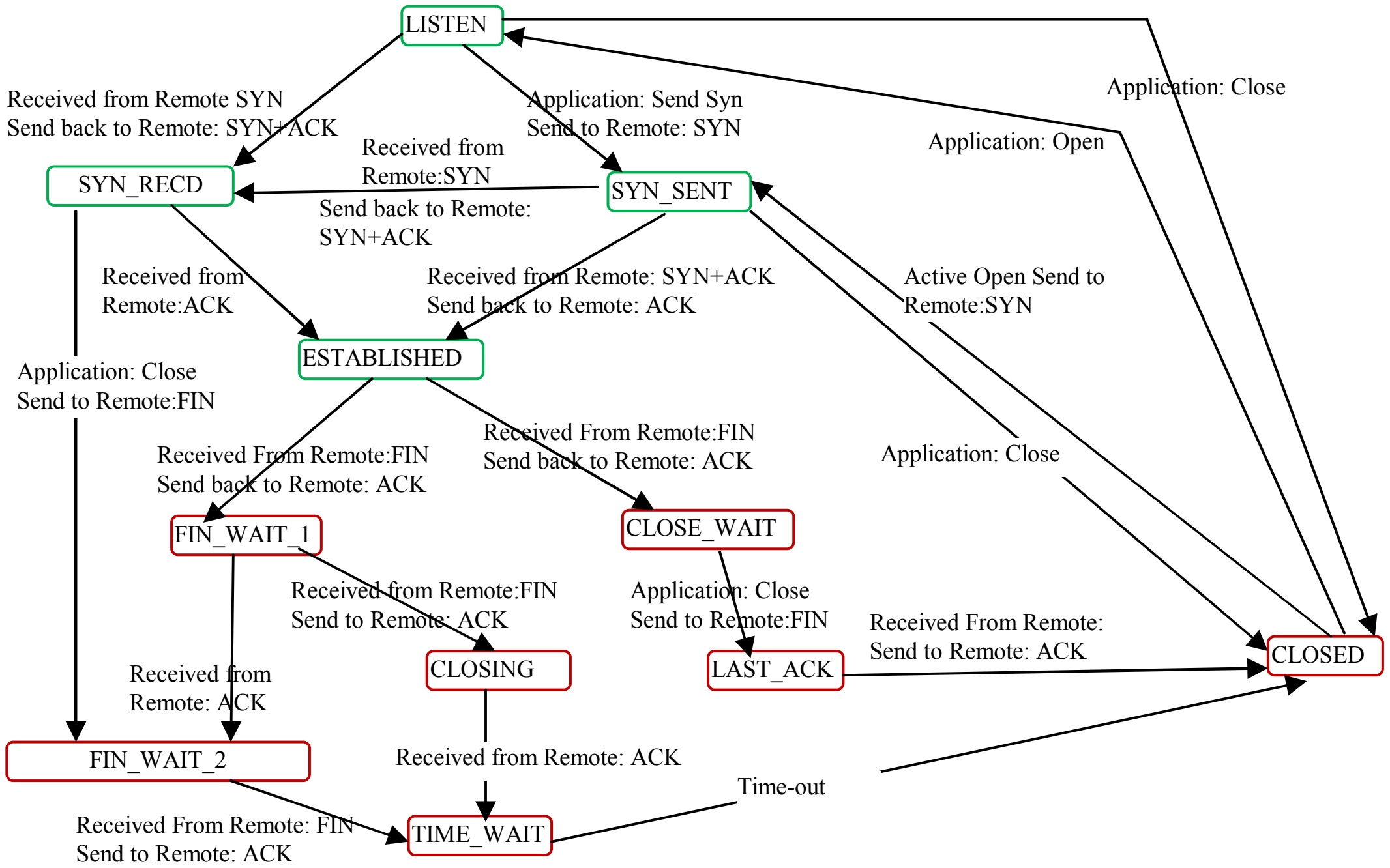

Fig. 6. TCP state transition diagram: The State of a TCP Connection at Local for a Connection between Local and Remote (From [15])

In Fig. 5, this is represented by the REMOTE (receiver) sending a Window (7) that is processed (8).

*If the receiver TCP sends 0 for the Window field, the sender TCP stops pushing segments into the IP layer on its side and starts what is known as the Persist Timer. When the Persist Timer expires, the sender TCP sends a small segment to the receiver TCP with the expectation that the ACK packet received in response will contain an updated value for the Window field.*
*It is in this state that data transfer takes place between the two end points.*

In Fig. 5, this depends on the processing (8) of the window. If the window is zero (9), this triggers the Persist Timer (10), and when the timer runs out (11), this, in turn, triggers the creation (12) of SYN that flows to REMOTE (13). REMOTE responds by sending ACK (14). Also, a Window of value 0 stops the manufacture of TCP segments (14). Notice that the TCP segment sphere has two flowsystems: a TCP segment and a State. Previously (e.g., in a data flowsystem), for simplicity's sake, we drew the sphere and the flowsystem in one rectangle, because the sphere had a single flowsystem. Also note that States are flowthings that can only be created and processed. In this case STOP (or whatever action occurs in the segment sphere) is a state that controls the activation/deactivation of the segment sphere.

Continuing from circle 12 in Fig. 5, the REMOTE sends ACK (15), triggering (16) activation of the sending of segments (17).

[At the beginning,] *When a local host wants to establish a connection with a remote host, it sends a SYN packet to the remote host. The remote should respond with a SYN+ACK packet, to which the local should send back an ACK packet. This is referred to as a three-way handshake: "SYN, SYN/ACK, ACK."*.

In Fig. 5, this three-way handshake is represented at circle 18, where processing of data triggers the creation of SYN that flows (19) to REMOTE. REMOTE then should send SYN+ACK (20), and this, upon receiving, triggers the creation of ACK (21) that flows to REMOTE (22).

The famous *three-way handshake: SYN, SYN/ACK, ACK* is typically represented by a message sequence diagram (Fig. 7) [16]. It can be represented in FM in an interesting way as shown in Fig. 8. The right side creates SYN (1) that flows to the left side (2), which processes it (3), triggering (4) creation of SYN+ACK. SYN+ACK flows to the left side (5), is processed, and triggers the creation of ACK (6). ACK flows to the left side (7). Because of the symmetry of the events, the three-way handshake appears as shown in Fig 9.

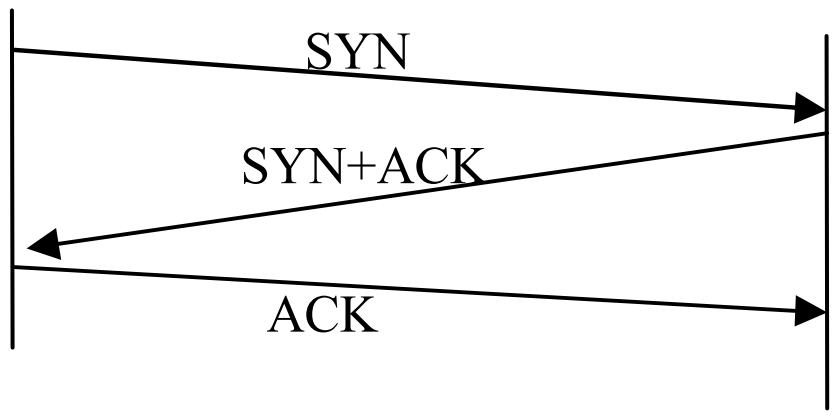

Fig. 7 Sequence diagram of a three-way handshake

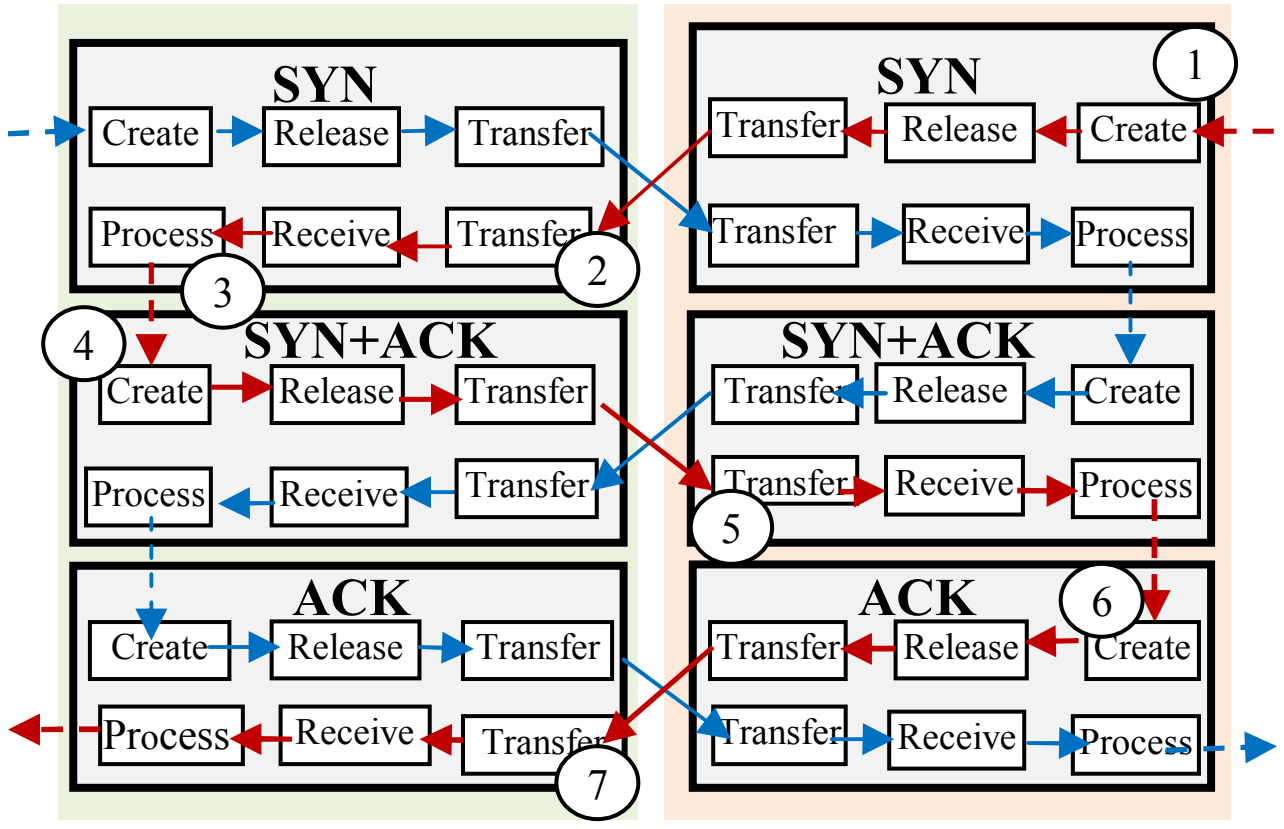

Fig. 8 The three-way handshake in FM

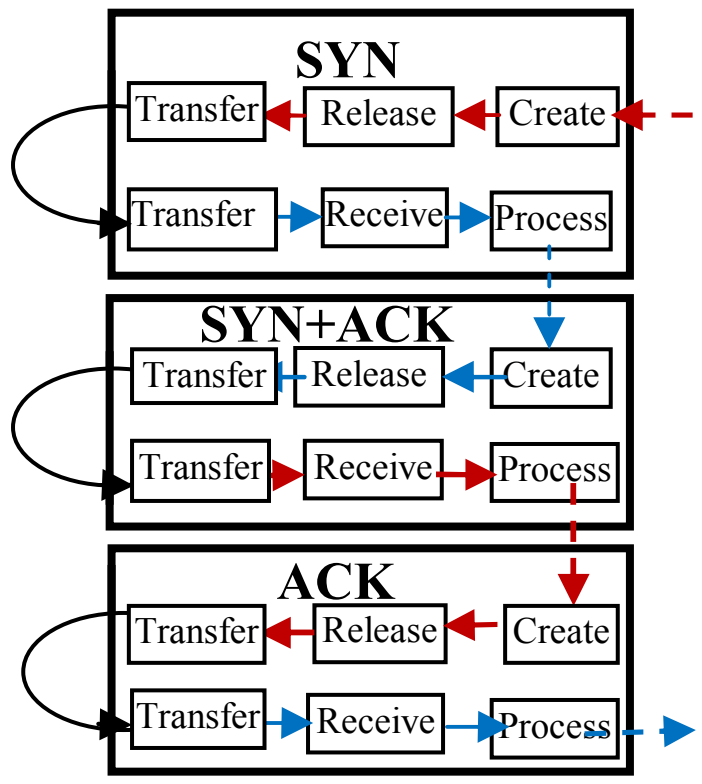

Fig. 9 Another representation of a three-way handshake

Returning to the discussion of Fig. 5, communication continues as follows:

*On the other hand, if the local host receives a SYN packet from a remote host, the local sends a SYN+ACK packet back to the remote. If the remote comes back with an ACK packet, the local transitions into the ESTABLISHED state. This again is a three-way handshake.*

In Fig. 5, this is represented at circle 23, where SYN is received, processed, and triggers the creation of SYN+ACK (24) that flows to REMOTE. If REMOTE comes back with ACK (15), transmission begins. Here, Kak [15] uses PI and REMOTE as the other side of the exchange. In Fig. 5, the transmission can be identified as receiving a TCP segment from IP (25).

*Regarding the state transition for the termination of a connection, each end must independently close its half of the connection.*
*If the local host wishes to terminate the connection first. It sends to the remote a FIN (Final—cleanly terminates a connection) packet and the TCP connection on the local transitions from ESTABLISHED to FIN WAIT 1. The remote must now respond with an ACK packet, which causes the local to transition to the FIN WAIT 2 state. Now the local waits to receive a FIN packet from the remote. When that happens, the local replies back with a ACK packet as it transitions into the TIME WAIT state. The only transition from this state is a timeout after two segment lifetimes to the state CLOSED.*

In the FM description, we distinguish between a request to terminate FIN (27) by the local host (the *client* that contacts the server), referred to as C-TERMINATE, and one issued by REMOTE (*Server*), referred to as S-TERMINATE. Accordingly, at circle 28, the local host creates C-TERMINATE (28) and sends (29) a request to terminate. Simultaneously, when sending the request it triggers a timing clock (30). IF REMOTE sends (31) FIN of type C-TERMINATE, the local host closes (32) the communication, triggering (33) sending (34) of ACK to REMOTE. If time runs out, the local host closes (35) the communication.

*When the remote host initiates termination of a connection by sending a FIN packet to the local. The local sends an ACK packet to the remote and transitions into the CLOSE WAIT state. It next sends a FIN packet to remote and transitions into the LAST ACK state. It now waits to receive an ACK packet from the remote.*

In Fig. 5, receipt of S-TERMINATE (36) by the local host triggers (37) sending (38) of ACK to REMOTE, which then creates and sends (39) FIN of type S-TERMINATE, simultaneously triggering a timing clock (40). If REMOTE then sends ACK (15), the local host closes (41) the communication; otherwise it closes (42) it when time has run out.

In this section, we have demonstrated how FM can describe TCP, with the aim of merely presenting the description as evidence of the viability of the methodology. We will now illustrate this viability in the area of security protocols.

## 6. Secure Sockets Layer

The Secure Sockets Layer (SSL) sits directly above the TCP to provide confidentiality and message integrity. The SSL security protocol is layered between the application protocol layer and the TCP/IP layer and can be divided into sublayers. We focus on the layers of the Handshake Protocol of the SSL protocol.

SSL takes the actual data to be sent, fragments it into blocks, applies authentication and encryption primitives to each block, and then sends the block to TCP for transmission over the network. On the receiving side, the blocks are decrypted, verified for integrity, reassembled, and delivered to the higher-level protocol. Before the SSL Record Protocol can do its thing, it must become aware of what algorithms to use for compression, authentication, and encryption. All of that information is generated by the SSL Handshake Protocol

To present a specific example, we concentrate on Microsoft's description [17] of a version of SSL protocol used in the Windows Server 2003 operating system. Its FM description is shown in Fig. 10. The architecture consists of the protocol suite that includes SSL.

> A client sends a message to a server, and the server responds with the information needed to authenticate itself. The client and server perform an additional exchange of session keys, and the authentication dialogue ends. When authentication is completed, secure communication can begin between the server and the client using the secret keys established during the authentication process [17].

The handshake protocol is a sequence of messages that negotiate the security parameters of a data transfer session. We follow the description given in [17] closely, summarizing, paraphrasing, and deleting irrelevant words and details as follows.

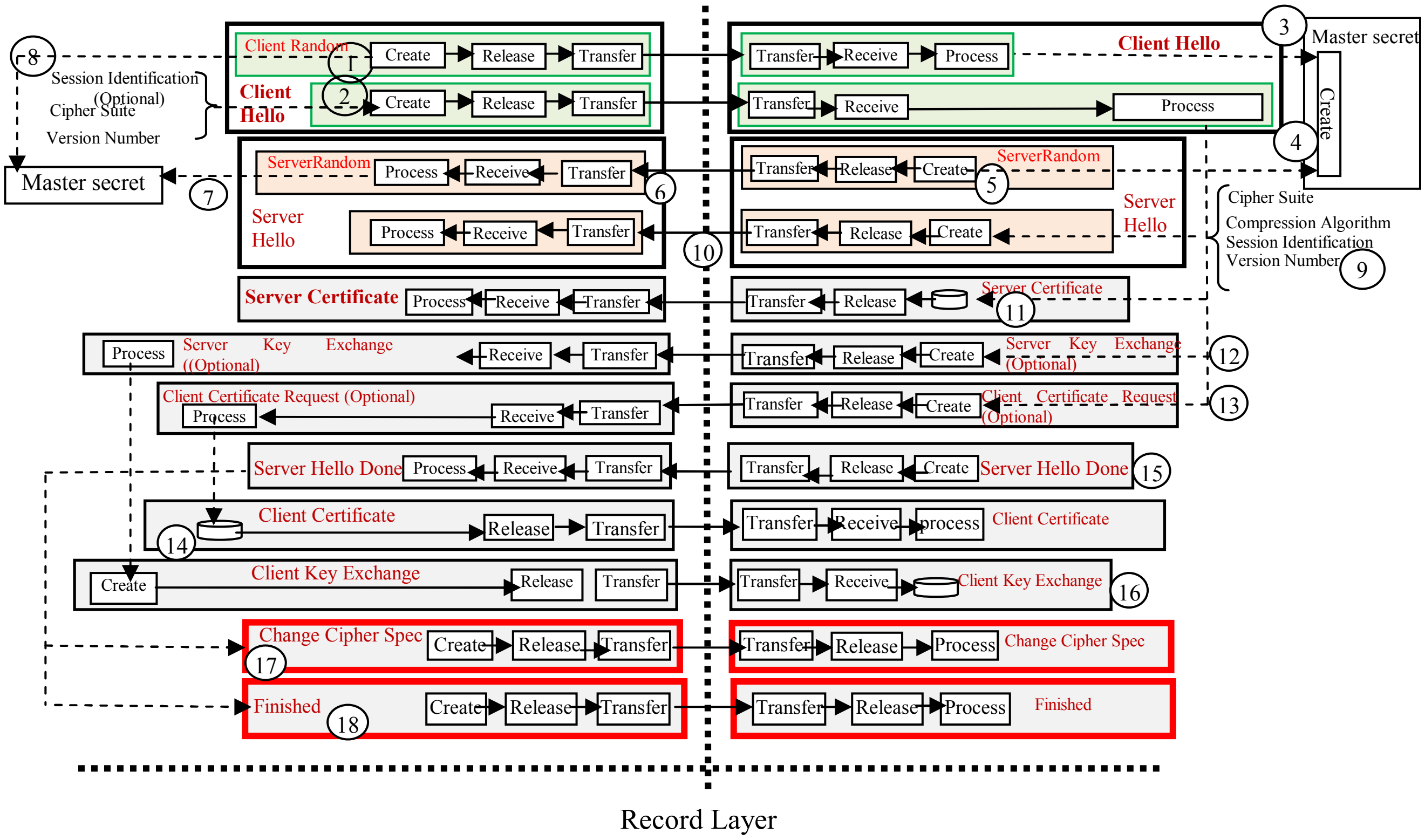

Fig. 10 FM representation of Microsoft's description [17] of the Handshake Protocol in a SSL protocol used in the Windows Server 2003 operating system, with some simplifications

*The client initiates a session by sending a Client Hello message to the server. The Client Hello message contains:*

- *Version Number*
- *(Optional) Session Identification*
- *Cipher Suite*
- *Client Random: A number that consists of the client's date and time, plus a cryptographically generated pseudorandom number. This is used in calculation of the Master Secret from which the encryption keys are derived.*

In Fig. 10, because of the special significance of the *Client Random*, we make two flowsystems, one for the Client number (circle 1) and the other for the remaining content of the Client Hello message (circle 2). The Client Random flows to the server, where it is used along with the Server Random in creating the Master Secret key (3 and 4).

*The server responds with a Server Hello message. The Server Hello message includes:*

- *Version Number*
- *Session Identification (if any)*
- *Cipher Suite*
- *Compression Algorithm, if used*
- *Server Random is a 4-byte representation of the server's date and time plus a 28-byte, cryptographically generated, pseudorandom number. This number, along with the Client Random, is used by both the client and the server to generate the Master Secret from which the encryption keys will be derived.*

In Fig. 10, the Server Random is created (5) and triggers creation of the master key (4). It also flows to the client (6), to be processed to create the master key (7) along with the Client Random (8). Also, the version number, session, identification (if any), cipher suite, and compression algorithm (9) are sent (10) to the client. These are represented as one message. Additionally,

- The Server Certificate is retrieved (11) and sent to the client.
- The Server Key Exchange (Optional) (12): "The server creates and sends a temporary key to the client. This key can be used by the client to encrypt the Client Key Exchange message later in the process. The step is only required when the server's certificate does not contain a public key that is suitable for key exchange" [17]. So (12) leads to (6), where the client encrypts and sends the key (16). Here we ignore representing the optional decision of this flow, even though the FM map can be detailed to track such details, e.g., checking if this requirement is needed when processing the server's certificate.
- The Client Certificate Request (Optional) (13) is sent to the client, which triggers (14) sending of the certificate.

These messages may trigger some other processes on the client side that can be described in FM. For example, "The server sends its certificate to the client. The server certificate contains the server's public key. The client *uses* this key to authenticate the server and to encrypt the Premaster Secret" [17] (italics added). Note that we make some simplifications (e.g., ignoring Certificate Verify Message) because the aim here is not to be very precise; rather the objective is to demonstrate the capabilities of the FM representation.

Lastly, the server sends the Server Hello Done message (16). Client responses to Server Hello include sending of the Client Certificate (14) and the Client Key Exchange (16), if required, as described previously. Also,

*The Change Cipher Spec message notifies the server that all future messages including the Client Finished message are encrypted using the keys and algorithms just negotiated… Both the client and the server have calculated the Master Secret. Up until now, however, any encryption has used the client's or server's private/public keys. The Change Cipher Spec message tells the server that the client is ready to use the Write Key for all further encryption.*

Accordingly, the *Change Cipher Spec* message is sent to the server (17). The *Finished message* is then sent by the client. It is the first message that the Record Layer encrypts.

## 7. Conclusions

This paper has presented a diagrammatic methodology for protocol specification that is applied to the Transmission Control Protocol and the Secure Sockets Layer. It is based on the notion of flow through six stages: creation, release, transfer, arrival, acceptance, and processing. The examples have demonstrated that the resultant conceptual description can provide a viable descriptive method for protocol specifications. Further research could experiment with applying the proposed representation to more specific protocols, with emphasis on protocol design, and applying the methodology in different networking areas such as wireless network [18, 20] and authentication [19].

**Sabah Al-Fedaghi** holds an MS and a PhD in computer science from the Department of Electrical Engineering and Computer Science, Northwestern University, Evanston, Illinois, and a BS in Engineering Science from Arizona State University, Tempe. He has published two books and more than 130 papers in journals and conferences on Software Engineering, Database Systems, Information Systems, Computer/information Ethics, Information Privacy, Information Security and Assurance, Information Warfare, Conceptual Modeling, System Modeling, Information Seeking, and Artificial Agents. He is an associate professor in the Computer Engineering Department, Kuwait University. He previously worked as a programmer at the Kuwait Oil Company and headed the Electrical and Computer Engineering Department (1991–1994) and the Computer Engineering Department (2000–2007). http://cpe.kuniv.edu/images/CVs/sabah.pdf